\documentclass[draft,jgrga]{agutex}

\usepackage{lineno}
\usepackage{amsmath}
\linenumbers*[1]

\usepackage[dvips]{graphicx}
\setkeys{Gin}{draft=false}
\authorrunninghead{STEFFL ET AL.}

\titlerunninghead{ENERGETIC ELECTRONS AT JUPITER}

\authoraddr{A.~J.~Steffl, A.~B.~Shinn, J.~Wm.~Parker, and S.~A.~Stern,
  Southwest Research Institute, 1050 Walnut St., Suite 300, Boulder, CO
  80302-5150, USA. (steffl@boulder.swri.edu)}

\authoraddr{G.~R.~Gladstone, K.~D.~Retherford, and M.~H.~Versteeg, Southwest
  Research Institute, P. O. Drawer 28510, San Antonio TX 78228-0510, USA.}

\begin{document}

%
%

\title{MeV electrons detected by the Alice UV spectrograph during the 
    New Horizons flyby of Jupiter}
%

%
%



\authors{A.~J.~Steffl\altaffilmark{1}, 
               A.~B.~Shinn\altaffilmark{1} , 
               G.~R.~Gladstone\altaffilmark{2}, 
               J.~Wm.~Parker\altaffilmark{1},
               K.~D.~Retherford\altaffilmark{2},
               D.~C.~Slater\altaffilmark{2,\dag},
               M.~H.~Versteeg\altaffilmark{2},
               S.~A.~Stern\altaffilmark{3}}

\altaffiltext{1}{Department of Space Studies, Southwest Research Institute,
  Boulder, Colorado, USA.}


\altaffiltext{2}{Space Science and Engineering Division, Southwest Research
  Institute, San Antonio, Texas, USA.}

\altaffiltext{3}{Space Science and Engineering Division, Southwest Research
  Institute, Boulder, Colorado, USA.}

\slugcomment{$^\dag$ Deceased 30 May 2011 }

%
%


\begin{abstract}
  In early 2007, the {\it New Horizons} spacecraft flew through the Jovian
  magnetosphere on the dusk side. Here, we present results from a novel means
  of detecting energetic electrons along {\it New Horizons}' trajectory: the
  background count rate of the Alice ultraviolet spectrograph. Electrons with
  energies $>$1~MeV can penetrate the thin aluminum housing of Alice, interact
  with the microchannel plate detector, and produce a count that is
  indistinguishable from an FUV photon. We present Alice data, proportional to
  the MeV electron flux, from an 11-day period centered on the spacecraft's
  closest approach to Jupiter, and compare it to electron data from the PEPSSI
  instrument. We find that a solar wind compression event passed over the
  spacecraft just prior to it entering the Jovian magnetosphere.
  Subsequently, the magnetopause boundary was detected at a distance of
  67~R$_J$ suggesting a compressed magnetospheric configuration. Three days
  later, when the spacecraft was 35-90~R$_J$ downstream of Jupiter, {\it New
    Horizons} observed a series of 15 current sheet crossings, all of which
  occurred significantly northward of model predictions implying solar wind
  influence over the middle and outer Jovian magnetosphere, even to radial
  distances as small as $\sim$35~R$_J$. In addition, we find the Jovian
  current sheet, which had a half-thickness of at least 7.4~R$_J$ between 1930
  and 2100~LT abruptly thinned to a thickness of $\sim$3.4~R$_J$ around
  2200~LT.
\end{abstract}

%
%

%

\begin{article}

%
%

\section{Introduction}

\begin{figure}
\noindent\includegraphics[width=20pc]{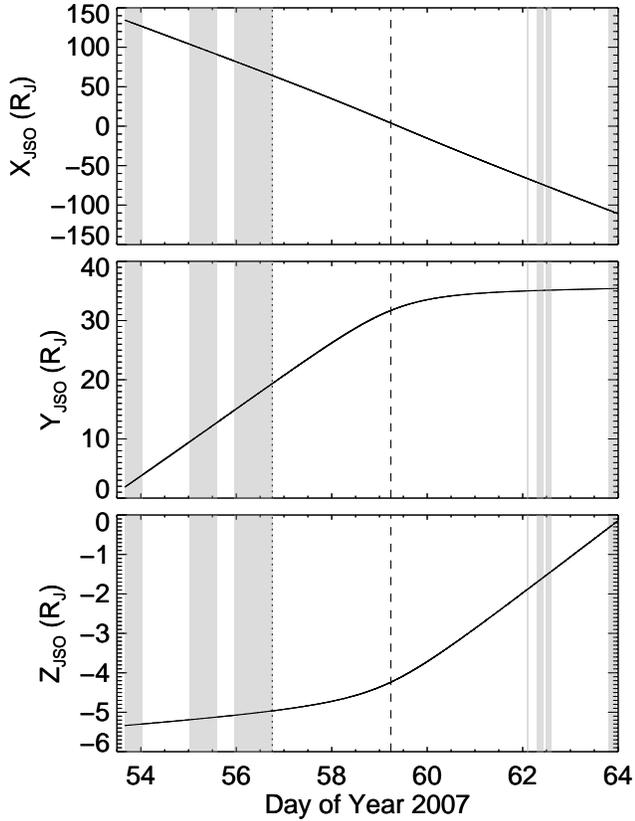}
\caption{\label{nh_jso_fig} Trajectory of the {\it New Horizons} spacecraft
  during the Alice observing period.  Distances are jovicentric and given in
  the Jupiter-Solar-Orbital coordinate system (+X to the Sun, +Z perpendicular
  to the orbit plane of Jupiter near the north rotational pole, and +Y
  completing the triad in the direction roughly opposite of Jupiter's orbital
  motion). The magnetopause crossing is marked by a dotted vertical line and
  the closest approach to Jupiter is marked with a dashed line. Shaded areas
  indicate periods when Alice was taking data and the energetic electron flux
  is consistent with {\it New Horizons} being on open magnetic field
  lines. 1~R$_J$ = 71,492 km}
\end{figure}

The {\it New Horizons} spacecraft, the first of NASA's New Frontiers program,
will be the first to flyby the Pluto system \citep{Sternetal08NHoverview}. In
early 2007, {\it New Horizons} flew past Jupiter on a gravity assist
trajectory that will bring it to its closest approach with Pluto on 14 July
2015. This flyby provided a unique opportunity to study the Jovian
magnetosphere and its response to a large increase in the dynamic pressure of
the solar wind. The spacecraft approached Jupiter from the direction of late
morning local time. {\it New Horizons}' closest approach to Jupiter occurred
at 05:43:41 UTC on 28 February 2007 (DOY 059) on the dusk side of Jupiter at a
distance of 32.2~R$_J$. The outbound trajectory was nearly aligned with the
anti-solar direction, which enabled NH to fly down the Jovian magnetotail out
to distances greater than 2500 R$_J$ \citep{Mccomasetal07, Mcnuttetal07}. The
trajectory of {\it New Horizons} around closest approach to Jupiter is shown
in Fig.~\ref{nh_jso_fig}.

The Alice FUV spectrograph on {\it New Horizons} consists of a small telescope
(4~cm x 4~cm) with an opaque aperture door that feeds a Rowland circle
spectrograph with a microchannel plate (MCP) detector and the associated
electronics \citep{Sternetal08}. These components are housed inside an
aluminum case just 1.3~mm (50~mils) thick.  During the Jupiter flyby, the
Alice FUV spectrograph made numerous observations of Jupiter, the Galilean
satellites, and the Io plasma torus over an 11-day period from day
053--064. After day 064, the angle between Jupiter and the Sun, as seen from
{\it New Horizons}, dropped below 20$^{\circ}$, precluding further FUV
observations. Results from Alice FUV observations of the Jovian system have
been presented by \cite{Gladstoneetal07}, \cite{Retherfordetal07}, and
\cite{Greathouseetal10}.

Alice is mounted to an exterior panel of the {\it New Horizons} spacecraft,
where it is directly exposed to the local energetic particle environment.
Incident particles from nearly 2$\pi$ steradians (the spacecraft body provides
shielding for the other 2$\pi$ steradians) can strike the instrument and, if
they have energies greater than 680~keV for electrons or 15~MeV for protons,
penetrate the relatively thin instrument housing. At Jupiter, the flux of
electrons with energy greater than 680~keV is several orders of magnitude
greater than the flux of protons with energy greater than 15~MeV
\citep{Baker:vanallen76, Schardtetal81}, so the penetrating energetic
particles will be dominated by electrons. Since the energetic particle flux at
Pluto is expected to be low, no additional attempt was made to shield the
detector. As a result, some fraction of the penetrating energetic particles
(or the secondary gamma-rays/x-rays they produce) are scattered in such a way
that they interact with the MCP detector to produce a count that is
indistinguishable from one made by a FUV photon. In this way, Alice can act as
a third {\it in situ} particle detector and complement the two dedicated
particles instruments on {\it New Horizons}, PEPSSI and SWAP. During the
Jupiter flyby, Alice demonstrated that it is quite effective at detecting
energetic electrons--at times generating more than 15,000 counts~s$^{-1}$.

PEPSSI, the Pluto Energetic Particle Spectrometer Science Investigation, is a
multi-directional, time-of-flight spectrometer that measures energetic
electrons and ions \citep{Mcnuttetal08}. We use PEPSSI electron data for
comparison to the Alice energetic particles data and during times when Alice
data are not available. PEPSSI contains three separate electron detectors,
each with a field of view of approximately $12.5^\circ \times 12^\circ$,
oriented in three separate look directions. Each detector measures electron
fluxes in three energy ranges: 25--190~keV, 190--700~keV, and 700--1000~keV.
Although PEPSSI can operate with an integration period as short as 1s, during
the Jupiter flyby, the integration period was generally 60s, although there
were several short periods where a 15s integration period was used.

\subsection{Jovian Coordinate Systems \label {coords_section}}

In this paper we use two different body-centered, non-inertial coordinate
systems: cylindrical System III (1965) and Cartesian Jupiter-Sun-Orbital
(JSO). System III (1965), usually referred to as just ``System III'', is a
coordinate system that rotates with (or nearly with) Jupiter. It is defined
such that the +Z axis is aligned with Jupiter's north rotational pole, while
the X and Y axes rotate with the planet at a rate of 870.536$^{\circ}$ per
ephemeris day \citep{Riddle:warwick76}. By tradition, System III is a
left-handed system, such that the longitude of an observer fixed in inertial
space increases with time.

JSO is a Cartesian coordinate system defined such that the +X axis is the
instantaneous vector from Jupiter to the Sun and the Z axis is perpendicular
to Jupiter's orbital plane, with +Z pointing close to Jupiter's north pole of
rotation. The Y axis is perpendicular to the other two axes to complete the
triad, with positive Y pointing approximately in the direction opposite of
Jupiter's orbital motion. To minimize confusion, an uppercase ``Z'' will refer
to the System III coordinate system while a lowercase ``z'' will refer to the
JSO coordinate system.

\section{Sensitivity of Alice to energetic particles}
\label{sensitivity_section}

There are two primary mechanisms through which Alice can detect energetic
particles: direct impingement on the MCP and FUV fluorescence induced in the
MgF$_2$ window located just in front of, and at 90$^\circ$ to, the front
surface of the MCP (cf. Fig.~5 in \cite{Sternetal08}). To quantify the
detection efficiency of these two mechanisms, testing was performed on the
flight spare detector of the {\it Rosetta} Alice spectrograph (the flight
spare detector for {\it New Horizons} Alice was used as the flight detector
for the LAMP instrument on the Lunar Reconnaissance Orbiter) at the Goddard
Space Flight Center Radiation Effects Facility (REF) in September 2003. The
{\it Rosetta} Alice flight spare detector is very similar to the {\it New
  Horizons} Alice flight detector and the input surface of the {\it Rosetta}
Alice MCP was re-coated using the same KBr and CsI photocathodes and layout as
used with {\it New Horizons} Alice.  The spare detector was placed in the REF
beam line and irradiated with 1~MeV electrons. The detection efficiency was
found to be 0.33 for direct irradiation of the MCP with MeV electrons and
0.012 for irradiation of the MgF$_2$ window. In addition, it was
confirmed that the detector is sensitive to gamma-rays produced by the
REF. Although not directly measured for this detector, similar MCP detectors
have quantum efficiencies of approximately 2\% for X-rays/gamma-rays with
energies between 50-2000~keV (O. Siegmund, personal communication, 2003).

The count rate of the detector was linear with the flux of particles.  In
addition, the observed variance of the Alice data matches the theoretical
prediction for Poisson-distributed events and a detector with a
non-paralyzable dead time \citep{Lucke76}. This strongly implies (but does not
strictly prove) that when an energetic electron interacts with the detector,
it produces a single count, rather than multiple counts, which would be
correlated.

\begin{figure}
\noindent\includegraphics[width=20pc]{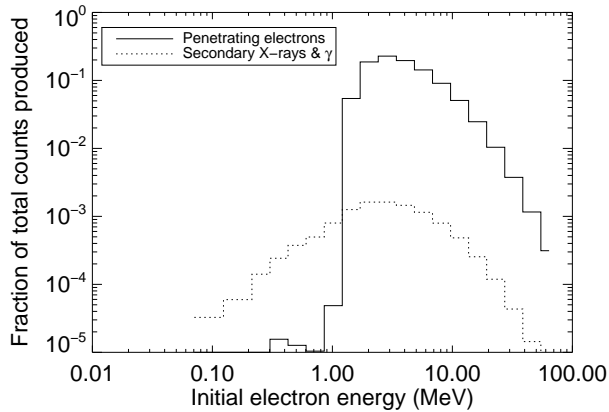}
\caption{\label{spenvis_fig} Fraction of the total number of Alice counts
  produced by penetrating electrons and secondary X-rays and $\gamma$-rays as
  a function of initial electron energy. The electrons are assumed to have an
  energy spectrum like that observed in the Jovian magnetosphere by {\it
    Pioneer 10} \citep{Baker:vanallen76}. This analysis does not include the
  effect of fluoresence from the window in the detector door. The efficiency
  of the Alice detector is taken to be 0.33 for electrons and 0.02 for X-rays
  and $\gamma$-rays (see text in Section~\ref{sensitivity_section} for
  details). 90\% of detected counts are produced by electrons with initial
  energies between 1-8~MeV. Electrons with initial energies as low as 50~keV
  can produce secondary photons that can be detected, but their effect on the
  total count rate is negligible.}
\end{figure}

We used the the MUlti-LAyered Shielding SImulation Software (MULASSIS)
\citep{Leietal2002} as implemented by the Space Environment Information System
(SPENVIS) \citep{Heynderickxetal2001} to model the flux of penetrating
electrons and secondary photons (X-rays and gamma-rays) through the interior
wall of the instrument housing as a function of initial electron energy over
the range of 0.05-64~MeV in 20 logarithmically spaced steps. By multiplying the
flux of penetrating electrons and secondary photons at each energy level by
the electron flux observed in the middle/outer magnetosphere of Jupiter
\citep{Baker:vanallen76} and the above detection efficiencies, we obtain an
estimate of the energy range of electrons most likely to produce Alice counts
at Jupiter, shown in Fig.~\ref{spenvis_fig}. 91\% of the Alice counts result
from penetrating electrons with initial energies between 1--8~MeV, with the
most probable energy range being 2-2.8~MeV. Electrons with initial energies
less than 1~MeV accounted for a scant 0.2\% of the total simulated
counts. Secondary photons produced by electrons of all initial energies
accounted for just 1\% of the total Alice counts, but 96\% of the counts
produced by electrons with initial energies less than 1~MeV.

\section{Alice Data and Analysis}

During the Jupiter flyby, the Alice FUV spectrograph made observations of
Jupiter, the Galilean satellites, and the Io plasma torus over an 11-day
period from day 053--064. After day 064, the angle between Jupiter and the
Sun, as seen from {\it New Horizons}, dropped below 20$^{\circ}$, precluding
further FUV observations. Results from Alice FUV observations of the Jovian
system have been presented by \cite{Gladstoneetal07}, \cite{Retherfordetal07},
and \cite{Greathouseetal10}.

Typically, Alice FUV observations lasted about one hour and were separated by
several hours.  To avoid repeated cycling of the instrument's high voltage
power supply between planned observations, Alice was often left powered-on at
the nominal high voltage level but with its opaque aperture door closed. In
this state, Alice was sensitive to photons (or energetic particles), but
external FUV photons could not reach the detector. In total, Alice spent
37.0\% of the time between DOY~053.7-064.0 like this.

Whenever Alice is on, it records housekeeping (HK) data at a programmable
rate. During the Jupiter flyby, this rate was once per second.  In addition to
information about the health of the instrument and its current operating
state, the HK data contain the total detector count rate. Although lacking any
spatial or spectral information, this data allows the instrument to act as a
simple photometer, even when it is not taking actively commanded exposures.

As a safety measure, if the count rate recorded in the Alice HK data ever
exceeds a programmable value, the instrument will end any currently active
exposure, turn off the detector high voltage, and enter ``safe'' mode for a
period of 15 minutes. After the safety timeout expires, the detector high
voltage remains off and Alice is incapable of detecting photons or energetic
particles until it receives a command to make an exposure or otherwise ramp up
the high voltage to operational levels. At the start of the Jupiter flyby
period, the Alice count rate safety value was set to 15,000 counts
s$^{-1}$. When {\it New Horizons} first entered the Jovian magnetosphere,
Alice repeatedly exceeded this count rate limit, resulting in a significant
loss of data on days 057--059. At 2007-059T18:01:17, roughly 12 hours after
closest approach, the Alice count rate limit was raised to 35,000 counts
s$^{-1}$. However, the observed count rate did not exceed the original limit
of 15,000 counts s$^{-1}$ (hereafter cps) for the remainder of the flyby.

\subsection{Detector Dead Time}
\label{deadtime_section}

When a charge cloud from the MCP hits the Alice double delay line (DDL)
readout anode, the detector electronics take a small, but non-zero, amount of
time to process the event. During this time, the detector cannot process any
additional events, i.e., it is ``dead''. This dead time introduces a
non-linearity to the detector response which becomes more significant at
higher count rates. Since the Alice detector operates in the non-paralyzable
regime the relationship between the true count rate and the observed count
rate is:

\begin{equation}
\label{deadtime_eq}
C_{true} = \frac{C_{obs}}{1 - \tau C_{obs}}
\end{equation}

\noindent where $\tau$ is time constant of the electronics. (See
\cite{Knoll1979} for further discussion of detector dead time and paralyzable
versus non-paralyzable dead time). The time constant of the Alice HK
electronics was measured in the lab and in-flight to be 4~$\mu$s. Thus, the
maximum dead time correction factor during the Jupiter flyby is 1.064, for an
observed HK count rate of 15,000~cps.

\subsection{Detector Background Count Rate}

Whenever the Alice detector is in its nominal operating state, it produces
counts at a low rate--even if the aperture door is closed and no FUV photons
can reach the detector. Primarily, these background counts are caused by gamma
rays from the spacecraft's RTGs striking the detector, although radioactive
decay in the MCP glass and noise in the detector electronics also
contribute. The intrinsic background count rate of Alice was measured in
August 2006, January 2007, and July 2007 and found to be constant, to within
5\%, with an average value of 104~counts~s$^{-1}$. After correcting the Alice
count rate data for detector dead time, we subtract this value from the data.

In addition to the intrinsic background count rate, the detector electronics
can be commanded to produce artificial events or ``stims pulses'', at fixed
locations on opposite ends of the readout anode. These stim pulses aid in
determining the location of where charge pulses from the MCP strike the
readout anode. When they are on, the stim pulses produce counts at a constant
rate of 36~counts~s$^{-1}$, which we subtract from the count rate data.

\subsection{FUV photons \label{fuv_correction}}

When the Alice aperture door is closed, all of the detected counts are due to
either the intrinsic background count rate (which only varies on timescales of
months or longer) or interactions with energetic particles. The situation
becomes more complicated when the aperture door is open and several thousand
counts per second are produced by FUV photons. We exclude from our analysis
all periods when Alice is observing a target that is known to vary rapidly
with time (e.g., the Jovian aurora or the Io plasma torus) or when the
spacecraft pointing is not constant during an observation, as in a scan. For
the remaining periods when the Alice aperture door is open, we estimate the
count rate due to FUV photons by comparing the mean count rate during the 15
seconds immediately prior to opening the aperture door to the mean count rate
in the 15 seconds immediately after. We attribute the increase in count rate
when the door is opened to be due solely to FUV photons. Likewise, we compare
the mean count rate during the 15 seconds immediately before and after the
aperture door was closed. We require the FUV photon count rate derived at the
beginning of the door open period to be within 25\% of the photon count rate
at the end of the door open period. If this condition is met, we assume that
any variations in the FUV photon count rate with time are small enough that
they can be approximated by a linear fit and then subtract off this fitted
count rate from the dead time corrected data. As seen in
Figs.~\ref{cntrate_all_fig}, \ref{mpcross_fig}, and \ref{evening_cntrate_fig}
below, count rate data obtained with the aperture door open and corrected in
this way closely matches the count rate data obtained with the aperture door
closed.

\section{Why use an FUV spectrograph to measure MeV electrons?}

Alice was clearly not designed to serve as an energetic electron detector. It
has no means of determining the spatial distribution of energetic electrons
nor does it have any capability to measure their energy, save for the low
energy cutoff required to penetrate the instrument housing. Since the
efficiency of detecting energetic electrons depends on both of these, there is
no way to reliably convert the Alice count rates into units differential
flux--although the count rate will be proportional to the integrated flux.
Given these rather significant drawbacks and the presence of a dedicated
energetic electron instrument onboard {\it New Horizons}, why is the Alice
electron data worth examining?

There are several reasons. Chief among these is that Alice is the only
instrument on {\it New Horizons} sensitive to electrons with energies greater
than 1~MeV. (Since electrons with these energies are not expected at Pluto,
PEPSSI was not designed to measure them.) Thus, Alice provides electron
measurements in a complimentary energy range. Second, there are several
periods during the 11-day period around closest approach when Alice was
collecting data but PEPSSI was either not powered on or was operating at
off-nominal instrument settings. Third, {\it New Horizons} was operated in
3-axis stabilized mode until DOY~82. With three electron detectors that have a
12.5$^\circ$x12$^\circ$ field of view, PEPSSI sampled only a small fraction of
the total 4$\pi$ steradians. Thus, PEPSSI could easily miss (or undercount) a
source of energetic electrons that is collimated and not omnidirectional. If
there are such collimated electrons, the PEPSSI count rates will be quite
sensitive to changes in spacecraft pointing, and indeed, at times there are
strong correlations between PEPSSI count rates and spacecraft attitude. The
roughly hemispheric coverage of Alice makes it less likely to miss a
collimated source, and less sensitive to small changes in pointing. Fourth,
the Alice count rates from MeV electrons are 100--300 times greater than the
count rates in the highest energy bin of the PEPSSI electron detectors,
resulting in a higher signal-to-noise ratio (due to Poisson noise). Alice data
also have 16 bits of linear dynamic range, whereas PEPSSI electron data have
10 bits of logarithmically scaled dynamic range. This results in the number of
electron events recorded in each PEPSSI integration period being quantized by
the six most significant bits needed to represent a given integer value. For
example, the following are sequences of three consecutive event counts that
PEPSSI can store: [61, 62, 63], [64, 66, 68], [128, 132, 136], [1024, 1056,
1088], etc. This introduced an uncertainty of up to 3\% in the true count
rate. Finally, the Alice data are sampled with 1s time resolution, compared to
60s time resolution for most of the PEPSSI electron data at Jupiter.

\section{Results}

Over a 11-day period, beginning on DOY 053 and roughly centered on the
spacecraft's closest approach to Jupiter, Alice recorded 328,954 measurements
of the count rate with the aperture door closed. An additional 61,977
measurements were recorded when the aperture door was open and the correction
for FUV photons could be applied. These measurements cover 44\% of the total
elapsed time during this period. The dead time corrected, background
subtracted Alice HK count rate during the Jupiter encounter is shown in
Fig. \ref{cntrate_all_fig}.  Data points plotted in red represent times when
the Alice aperture door was open and the FUV photon count rate has been
subtracted as described in Section~\ref{fuv_correction}, above.  Since, in all
cases, the FUV count rate subtracted from data obtained with the aperture door
open was several thousand counts s$^{-1}$, these data have a much larger
variance than data obtained nearby in time, but with the aperture door closed.

\subsection{Upstream Solar Wind Conditions}
\label{upstream_section}

Prior to crossing the Jovian magnetopause, the {\it New Horizons} spacecraft
was immersed in the ambient solar wind plasma. Alice remained off during the
distant approach to Jupiter until DOY 053. However, PEPSSI did take data
during this period, and from DOY~040, at a distance of 442~R$_J$ upstream
Jupiter, until DOY~053, 134~R$_J$ upstream Jupiter, the electron detectors on
PEPSSI show a gradual 15\% increase in the flux of energetic electrons. Given
the long duration of this increase and correlation with distance, the only
plausible source of these electrons is the Jovian magnetosphere. A similar
enhancement of energetic particles was seen by the {\it Voyager} spacecraft on
approach to Jupiter \citep{Bakeretal1984}. When Alice started taking data on
DOY~053 at 16:00, the average count rate was 10~cps higher than the intrinsic
background rate of 104~cps, indicating the presence of a detectable flux of
Jovian energetic electrons.

Results from the Michigan Solar Wind Model (mSWiM) which uses measurements at
Earth to propagate solar wind conditions to Jupiter using a 1-D MHD code
\citep{Zieger:hansen08}, predicted that a large solar wind compression event
would reach Jupiter on DOY~051$\pm$1. Two days later, on DOY~053, the Solar
Wind Around Pluto (SWAP) instrument on New Horizons \citep{McComasetal08},
observed a solar wind forward shock followed by an increase in solar wind
protons, consistent with the arrival of this event \citep{Elliott2007mop}.
This solar wind event seems responsible for the sharp increase in the
brightness of the Jovian FUV aurora observed by HST on DOY 054 and 055
\citep{Clarkeetal09}. After adjusting the propagated arrival times by 2.1 days
to match the SWAP observation of the solar wind forward shock, mSWiM predicts
a sharp rise in solar wind density at the position of {\it New Horizons}
beginning at DOY~054.8 with a peak solar wind density occurring at
DOY~055.1. PEPSSI was taking data throughout this period, and beginning on
DOY~054.9, each of the PEPSSI's three electron detectors measured a
significant increase in the flux of energetic electrons. 14 hours of Alice
count rate data are also available during the period of peak solar wind
density, as shown in Fig.~\ref{upstream_fig}. The ratio of electron flux at
the peak of the solar wind event to the pre-event background rate is 9.4 for
Alice and 4.0, 3.3, and 3.1 for the high, medium, and low energy ranges of the
PEPSSI E0 detector, implying that the electron energy spectrum in the solar
wind flow was significantly harder than in the ambient plasma consisting of
electrons coming from Jupiter.

The Alice count rate during this period is highly correlated with the observed
energetic electron fluxes from all three of the PEPSSI electron detectors. The
correlation is strongest for the PEPSSI E0 detector. Geometrically, this makes
sense, since the E0 detector has it's field of view in the same hemisphere of
the sky as Alice (the +Z hemisphere, in spacecraft coordinates). The PEPSSI E1
and E2 detectors both have their boresight vectors at an angle of -22$^\circ$
relative to the spacecraft X-Y plane, and electrons from this hemisphere would
have to pass through the spacecraft body before reaching Alice. The Pearson
linear correlation coefficient between Alice and the PEPSSI E0 data is 0.88
for 25-190~keV electrons, 0.95 for 190-700~keV electrons, and 0.86 for
700-1000~keV electrons bin. In this case, the low correlation coefficient for
the highest PEPSSI energy range is likely due to the relatively low number of
electrons detected by PEPSSI (10-100 per integration period); the correlation
increases when the PEPSSI data are binned in time to increase their
signal-to-noise.

\subsection{Upstream Magnetopause Crossing}

At 17:50 UTC on DOY 056, at a distance of 67~R$_J$ from Jupiter, the Alice
background subtracted count rate increased from 7~cps to 7000~cps in 80
minutes. This sudden increase in count rate, shown in Fig.~\ref{mpcross_fig},
is due to the increase in the energetic electron flux as {\it New Horizons}
crossed the Jovian magnetopause. After this sharp initial increase, the count
rate decreased for the next two hours, reaching a level of 2800~cps, before
increasing again to a dead time corrected count rate of 15800~cps at which
point the instrument was repeatedly tripped into safe mode producing a gap in
the count rate data. The three electron detectors on PEPSSI all show similar
behavior. The ratios of the Alice count rate and PEPSSI electron fluxes before
and after the magnetopause crossings show a harder electron energy spectrum
inside Jupiter's magnetosphere. Again, there is a strong correlation between
the Alice count rate and the PEPSSI E0 electron flux, with Pearson linear
correlation coefficients of 0.89 for the 25-190~keV energy range, 0.89 for the
190-700~keV energy range, and 0.94 for the 700-1000~keV energy range.

The location of the Jovian magnetopause boundary is determined by the balance
between the dynamic pressure of the solar wind and the internal pressure of
the magnetosphere, which consists of dynamic, thermal, and magnetic components
\citep{Huddlestonetal98}. \cite{Joyetal02} performed a statistical analysis on
all magnetopause crossings observed by the {\it Pioneer 10}, {\it Pioneer 11},
{\it Voyager 1}, {\it Voyager 2}, {\it Ulysses}, and {\it Galileo} spacecraft
and found that the Jovian magnetopause boundary exhibits a bimodal
distribution with the most probable standoff distances occurring at
92$\pm$6~R$_J$ when the magnetosphere was in an expanded state and
63$\pm$4~R$_J$ when the magnetosphere was in a compressed state. The observed
magnetopause location at 67~R$_J$ implies that the magnetosphere was in a
compressed configuration, as might be expected given the observation of the
prior solar wind event.

The PEPSSI E0 electron fluxes also show dramatic spikes prior to the
magnetopause crossing around DOY~056.25 and 056.35. \cite{Mcnuttetal07}
attributed this to either the Jovian bow shock crossing or to upstream events
like those reported by \cite{Haggerty:armstrong99}. Without data from a
magnetometer, it is difficult to distinguish between these two possible
scenarios. The distance of {\it New Horizons} from Jupiter during these two
spikes, 75--79~R$_J$, is consistent with the most probable bow shock distance
for a compressed magnetosphere: 73$\pm$10~R$_J$ \citep{Joyetal02}.  However,
the onset of the spike at DOY~056.25 corresponds exactly to when the
spacecraft was executing an 81$^\circ$ slew. There are no significant pointing
changes that can explain the return back to the pre-spike flux or the
subsequent spike at DOY~056.35. Furthermore, while the PEPSSI E0 detector high
energy electron flux shows nearly a factor of 30 increase in flux, the E1
detector shows only a 50\% increase and the E2 detector actually shows a 20\%
decrease. After the spacecraft slew, the boresights of the PEPSSI E0, E1, and
E2 detectors were oriented 64$^\circ$, 111$^\circ$, and 148$^\circ$ from the
{\it New Horizons}-Jupiter vector, respectively. This suggests that the flux
increases PEPSSI observed are due to bursts of electrons streaming away from
Jupiter. The lack of any detectable increase in the Alice count rate implies
that these bursts are limited to electrons with energies below 1~MeV.

\subsection{Dayside Magnetosphere}
As {\it New Horizons} flew through the dayside magnetosphere, the energetic
electron flux observed by Alice was quite intense.  Alice repeatedly exceeded
the count rate safety limit of 15000~counts~s$^{-1}$, producing the large data
gaps on DOY~057, 058, and 059 seen in Fig.~\ref{cntrate_all_fig}. When the
flux of energetic electrons was low enough to permit Alice to operate
normally, the observed count rate was highly variable on timescales of minutes
to hours, and there are numerous occurrences when the Alice count rate changes
by a factor of 2 on timescales of 50-100~s. Figure~\ref{dayside_cnt_fig} shows
the Alice count rate during a typical 30-minute period on DOY~057 beginning at
20:25:00 UT.

\subsection{Dusk Sector Current Sheet Crossings}

After {\it New Horizons}' closest approach to Jupiter, the flux of energetic
electrons was less intense. Figure~\ref{evening_cntrate_fig} shows the Alice
and PEPSSI E0 count rates from DOY~059.5--063, when {\it New Horizons} was in
the dusk-to-midnight sector. For comparison, the count rates have been
normalized to the average value between DOY~061.0--061.4.  Clear peaks in the
count rate were observed when {\it New Horizons} was near System III (1965)
longitudes of $\lambda_{III}$=130$^\circ$ and $\lambda_{III}$=280$^\circ$.
These peaks in count rate result from the center of the Jovian current sheet
sweeping over the spacecraft twice per Jovian rotation. The count rate peak
near $\lambda_{III}$=130$^\circ$ occurs when {\it New Horizons} crosses the
center of the current sheet from the south to the north (in reality, New
Horizons moves a relatively small distance north during one Jovian rotation
period, while the current sheet sweeps rapidly over the spacecraft in an
oscillatory manner). Likewise, the peak near $\lambda_{III}$=280$^\circ$
occurs when {\it New Horizons} crosses the center of the current sheet from
the north moving to the south, i.e., the current sheet is moving northward
over the spacecraft. Similar behavior has been seen by the {\it Pioneer 10},
{\it Voyager 1}, {\it Voyager 2}, and {\it Galileo} spacecraft when they were
near Jupiter's rotational equator \citep{Goertzetal76, Vogtetal79a,
  Vogtetal79b, Vasyliunasetal97, Waldropetal05}. In total, Alice observed 10
current sheet crossings. PEPSSI also observed 10 current sheet crossings, five
of which occurred during periods when Alice was not operating. The times of
the observed current sheet crossings and the location of the {\it New
  Horizons} spacecraft in System III (1965) and JSO coordinates are given in
Table~\ref{cs_table}.

\subsubsection{Location of Current Sheet Crossings}
\label{cs_xing_section}

The simplest model for Jupiter's current sheet is that of a rigid sheet lying
in the equator plane of the magnetic dipole field. The height of such a
current sheet is given by:

\begin{equation}
  \label{rigid_cs_eqn}
  Z_{rigid}=\rho \tan \theta \cos\left(\lambda - \lambda' \right)
\end{equation}

\noindent where $\rho$ is the cylindrical radial distance, $\theta$ is the
tilt of the magnetic dipole field, $\lambda$ is the System III longitude, and
$\lambda'$ is the System III longitude at which the current sheet has its most
northern (+Z) extent. For the VIP4 model of Jupiter's magnetic field,
$\theta=9.52^{\circ}$ and $\lambda'_{VIP4}=20.4^\circ$
\citep{Connerneyetal98}. Panel A in Figure~\ref{delta_z_fig} shows the
difference in height between {\it New Horizons} and a rigid current sheet as a
function of the spacecraft's (cylindrical) radial distance at the time of the
15 current sheet crossings given in Table~\ref{cs_table}. Filled (open)
symbols represent a crossing during which {\it New Horizons} went from being
north (south) of the current sheet to south (north) of it. The total RMS error
of the data points is 3.5~R$_J$ and the difference increases roughly linearly
with radial distance from Jupiter indicating that the actual current sheet
does not lie in the dipole equator plane, but rather is warped or hinged
northward.

A hinged current sheet lies in the magnetic dipole equator plane close to the
planet, but at larger distances is warped into a plane that is parallel to
either the rotational equator \citep{Behannonetal81, Khurana92} or the solar
wind flow \citep{Khurana:schwarzl05}. To date, the most accurate published
model of the Jovian current sheet is that of \cite{Khurana:schwarzl05},
hereafter KS05. Their model is based on fits to 6328 current sheet crossings
observed in magnetometer data from {\it Pioneer 10}, {\it Pioneer 11}, {\it
  Voyager 1}, {\it Voyager 2}, {\it Ulysses}, and {\it Galileo}. The height of
the hinged current sheet in the KS05 model is given by their Eq.~13:

\begin{eqnarray}
\label{khurana_z_eq}
Z_{model} & = & {\left(\sqrt{\left( x_H \tanh \frac{x}{x_H}\right)^2 +
      y^2} \right) \tan \theta} \cos\left(\lambda - \lambda' \right) \nonumber \\
&& + x \left( 1 - \tanh \left|
    \frac{x_H}{x} \right| \right) \tan \theta_{Sun}
\end{eqnarray}

\noindent where $x$ and $y$ are the JSO coordinates of the spacecraft,
$x_H=47$~R$_J$ is the characteristic hinging distance, and $\theta_{Sun}$ is
the angle between the Jupiter-Sun vector and the rotational equator. During
the {\it New Horizons} flyby, $\theta_{Sun}=-2.9^\circ$. KS05 also include the
effects of the non-dipolar magnetic field line geometry and the time to
propagate a signal from Jupiter to the current sheet along a magnetic field
line. As such, $\lambda'$ in Eq.~\ref{khurana_z_eq} is no longer a constant
and is instead given by

\begin{equation}
\label{lambda_prime_eq}
\lambda' = \lambda'_{VIP4} + \delta_{wave} + \delta_B + b_0
\end{equation}

\noindent where $\delta_{wave}$ is the shift of the prime meridian longitude
due to the finite time to propagate a magnetic signal from Jupiter, $\delta_B$
is the shift due to the swept-back geometry of Jupiter's magnetic field lines,
and $b_0$ is a constant offset term. Both $\delta_B$ and $\delta_{wave}$ are
complicated non-linear functions of local time and radial distance. We note
that for $\delta_B$, the value of the model parameter $a_4$ given in Table~1
of KS05 contains a typographical error. The correct value should be $a_4 =
0.0016$ (K. Khurana, personal communication, 2010).

Panel B of Fig.~\ref{delta_z_fig} shows the height difference between the
observed current sheet location and the KS05 model.  While the overall RMS
error of 2.3~R$_J$ is better than the simple rigid current sheet model (and
even somewhat better than the value of the RMS error of fit for all data used
by KS05), it is clear that the fit for N$\rightarrow$S crossings (RMS of
3.3~R$_J$) is systematically worse than that for S$\rightarrow$N crossings
(RMS of 0.8~R$_J$). This implies that one or more of the terms in
Eq.~\ref{lambda_prime_eq} require modification to better match the effective
prime meridian of the current sheet during the {\it New Horizons} flyby.
Since the data used to derive the parameters of the KS05 model were obtained
over a large range of local times, radial distances, and solar wind
conditions, this is not particularly surprising.

The full KS05 model is governed by 24 parameters. With only 15 current sheet
crossings observed along a single trajectory, the KS05 model is significantly
under-determined by the {\it New Horizons} current sheet data. We therefore
take the simplest approach of allowing the constant offset term in
Eq~\ref{lambda_prime_eq}, $b_0$, to vary to match the data. We find a value of
$b_0=12.0^\circ$ yields the best fit and results in a N$\rightarrow$S RMS of
2.4~R$_J$, a S$\rightarrow$N RMS of 2.0~R$_J$, and a combined RMS of
2.2~R$_J$, as shown in Fig.~\ref{delta_z_fig}, panel~C. Like the rigid current
sheet case shown in panel A, all of the observed current sheet crossings, even
those at 35~R$_J$, have been displaced northwards of the predicted
locations. Given that the Z-component of the solar wind velocity is also in
the northward direction, the most plausible explanation is that the current
sheet has been pushed northwards by the solar wind.

It has long been recognized that the hinging distance of the current sheet,
$x_H$ will vary in response to the solar wind dynamic pressure, moving outward
when the solar wind pressure is low and inward when the solar wind pressure is
high \citep{Goertz81}. Given that the mSWiM model predicts a factor of
$\sim$80 increase in the solar wind dynamic pressure at Jupiter, with elevated
levels remaining until DOY~062, it is quite plausible that the hinging
distance during the {\it New Horizons} flyby would be smaller than the value
of $x_H=47$~R$_J$ found by KS05. We therefore allow both $b_0$ and $x_h$ to
vary and find best-fit values of $b_0=12.5^\circ$ and $x_H=5.0$~R$_J$. This
combination of model parameters yields a combined RMS value of 0.6~R$_J$
N$\rightarrow$S, 0.5~R$_J$ S$\rightarrow$N, and 0.6~R$_J$ total--a noticeably
better fit, with no clear systematic deviations--as shown in panel D. (Note,
we strongly caution against extrapolating these fit results radially inwards
of the observed {\it New Horizons} current sheet crossings. There, strong
centrifugal forces will confine the current sheet to the ``centrifugal
equator'', the locus of points that are most distant from Jupiter's spin axis
along a given magnetic field line.) Thus, it seems highly likely that the
solar wind influences the location of the current sheet even to radial
distances as small as $\sim$35~R$_J$.

\subsubsection{Thickness of the Dusk Current Sheet}

Initially, the count rate peaks associated with current sheet crossings are
relatively broad (cf. Fig.~\ref{evening_cntrate_fig}). At the minima between
peaks, both Alice and PEPSSI count rates remain significantly elevated above
their pre-magnetopause crossing levels, with count rates only a few times
lower than their peak values.  This suggests that initially, at least, {\it
  New Horizons} did not exit the current sheet and enter into the lobe
regions, although from the low count rates on DOY~060.9 and DOY~061.3, it
appears that the spacecraft came close to entering the southern lobe
region. This implies that the current sheet in this sector of the dusk
magnetosphere is thicker than its vertical motion over the course of one
Jovian rotation.

Although we cannot directly measure the current sheet thickness with Alice or
PEPSSI data, we can estimate the height of the center of the current sheet
using the modified KS05 current sheet model described in
Section~\ref{cs_xing_section} above. Using the best-fit model parameters of
$x_H=5.0$~R$_J$ and $b_0=12.5^\circ$, the center of the current sheet was
7.4~R$_J$ north the spacecraft on DOY~060.9 (when {\it New Horizons} appears
to be closest to leaving the current sheet), and thus the half-thickness of
the current sheet at this location must be at least as great. This is
consistent with data from the {\it Ulysses} Energetic Particle Composition
Instrument and the {\it Galileo} magnetometer that also suggest a thick
current sheet in the dusk sector \citep{Kruppetal1999, Kivelson:khurana2002}.

However, a rapid transition in the current sheet appears to have occurred
around DOY~062, when the spacecraft was at a local time of $\sim$2200~LT and a
radial distance of $\sim$70~R$_J$. The count rate peak corresponding to the
current sheet crossing on DOY~061.86 is broad, similar to previous count rate
peaks, and indicative of a thick current sheet. However, the next four current
sheet crossings all have count rate peaks that are much narrower, and between
peaks, the count rates fall to the lowest levels seen inside the Jovian
magnetosphere (an Alice count rate of $\sim$150 cps). This low flux of
energetic electrons is consistent with {\it New Horizons} having entered the
lobe regions, where magnetic field lines are open and energetic particles are
quickly lost. Using the modified KS05 model, we estimate the current sheet
half thickness in this region to be $\sim$3.4~R$_J$--a value intermediate
between the $>$7.4~R$_J$ observed earlier and the half thickness of
$<$1.5~R$_J$ typical of the dawn sector.

There is some evidence for a similarly abrupt current sheet transition in {\it
  Galileo} magnetometer data. Beyond a radial distance of 30~R$_J$, the
magnetic pressure at the center of the was observed to decrease suddenly
around 2100~LT \citep{Kivelson:khurana2002}. One interpretation of this result
is that plasma has gone from being more distributed along the flux tube to
being more concentrated in the equatorial region, i.e., the current sheet has
gone from being thicker and relatively diffuse to thinner and more
dense. However, if the plasma densities were actually higher at the center of
the current sheet, one would expect the Alice and PEPSSI count rates to be
significantly higher than observed previously when the plasma was more
diffuse. Instead, the peak energetic electron fluxes are comparable to
previous maxima, suggesting that the plasma density at the center of the
current sheet hasn't changed significantly, though the sheet is thinner, which
in turn implies a lowered flux tube content. Intriguingly, the sudden thinning
of the current sheet correponds to the arrival of a solar wind reverse shock
(lower dynamic pressure following the shock) predicted by the mSWiM model to
arrive at Jupiter on DOY~061.9.

\section{Conclusions}

The Alice FUV spectrograph on {\it New Horizons} provides a novel means of
detecting 1-8 MeV electrons at Jupiter.  Taken as a whole, the Alice count
rate data, in combination with the PEPSSI electron data, show the response of
the Jovian magnetosphere to a period of disturbed solar wind flow. Fortuitous
timing and the high flyby velocity of the spacecraft enabled {\it New
  Horizons} to directly observe a solar wind compression event pass over the
spacecraft and subsequently fly through the Jovian magnetopause, the
outer-to-middle magnetosphere, and the dusk sector current sheet while the
solar wind dynamic pressure remained at elevated levels. The location of the
upstream magnetopause crossing at 67~R$_J$ (and possibly the bow shock
crossing at 75-79~R$_J$) is consistent with the magnetosphere being in a
compressed configuration. The northerly displacement of the 15 observed
current sheet crossings suggests the influence of the solar wind, even at
radial distances as small as 35~R$_J$. An abrupt thinning of the current sheet
was observed around 2200 LT (72~R$_J$), when the half-thickness of the current
sheet decreased from $>$7.4~R$_J$ to $\sim$3.4~R$_J$ over a few hours in
elapsed time, a few R$_J$ in radial distance, and a few minutes in local solar
time. We propose that this thinning is due to a temporal change, possibly
associated with the predicted arrival of a solar wind reverse shock.


\begin{acknowledgments}
  We would like to thank the {\it New Horizons} science and mission teams and
  acknowledge the financial support of the {\it New Horizons} mission by
  NASA. AJS and ABS gratefully acknowledge support from the NASA Jupiter Data
  Analysis Program (NNX09AE05G). We kindly thank Fran Bagenal, Peter Delamere,
  Mariel Desroche, Caitriona Jackman, and Krishan Khurana for many helpful
  discussions and K.C. Hansen and B. Zieger for providing solar wind
  propagations from their Michigan Solar Wind Model
  (http://mswim.engin.umich.edu/).
\end{acknowledgments}

%
%
%
%
%
%
%
%
%
%


%
%

\end{article}




%
%
%
%
%
%
\pagebreak
\begin{table*}
  \caption{\label{cs_table}Times and locations of {\it New Horizons} during current
    sheet crossings}
  \begin{tabular}{lcccccc}
    DOY & $\rho_{III}$ (R$_J$) & $\lambda_{III}$ ($^\circ$) & $Z_{III}$ (R$_J$) &
    x$_{JSO}$ (R$_J$) & y$_{JSO}$ (R$_J$) & z$_{JSO}$ (R$_J$) \\
     59.753$^\dag$ & 34.4 & 142.7 & -2.8 &   -9.1 &  33.1 &  -3.9 \\
     59.900$^\dag$ & 35.8 & 265.1 & -2.5 & -12.8 &  33.4 &  -3.8 \\
     60.179             & 39.3 & 138.4 & -1.9 & -19.8 &  33.8 &  -3.6 \\
     60.336             & 41.6 & 271.2 & -1.6 & -23.7 &  34.0 &  -3.4 \\
     60.590$^\dag$ & 45.6 & 125.5 & -1.1 & -30.0 &  34.2 &  -3.2 \\
     60.771             & 48.8 & 279.2 & -0.7 & -34.4 &  34.4 &  -3.1 \\
     61.027             & 53.5 & 137.9 & -0.1 & -40.7 &  34.6 &  -2.9 \\
     61.190             & 56.6 & 276.8 &   0.2 & -44.6 &  34.7 &  -2.7 \\
     61.431             & 61.3 & 123.7 &   0.7 & -50.5 &  34.8 &  -2.5 \\
     61.603             & 64.8 & 271.2 &   1.1 & -54.6 &  34.9 &  -2.3 \\
     61.861$^\dag$ & 70.1 & 133.3 &   1.7 & -60.8 &  34.9 &  -2.1 \\
     62.033             & 73.7 & 281.5 &   2.0 & -64.8 &  35.0 &  -1.9 \\
     62.269             & 78.7 & 125.1 &   2.5 & -70.4 &  35.1 &  -1.7 \\
     62.453             & 82.6 & 284.1 &   2.9 & -74.8 &  35.1 &  -1.6 \\
     62.686$^\dag$ & 87.6 & 125.3 &   3.4 & -80.3 &  35.2 &  -1.3 \\
    \hline
    \multicolumn{7}{l}{\footnotesize $^\dag$ Current sheet crossing derived
      from PEPSSI electron data}
    \end{tabular}
\end{table*}

\pagebreak

\begin{figure}
\noindent\includegraphics[width=39pc]{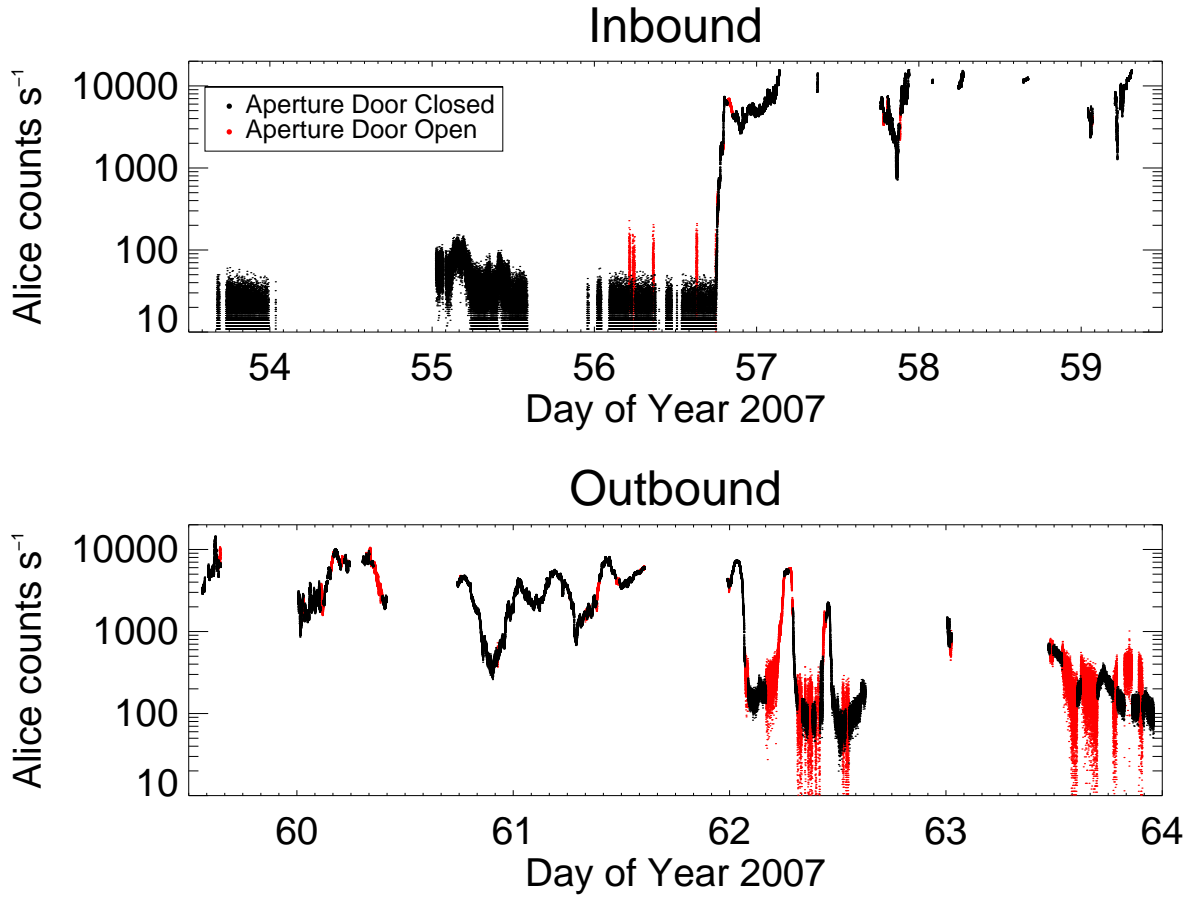}
\caption{\label{cntrate_all_fig} The count rate of Alice recorded in
  housekeeping data during the Jupiter flyby. The data have been corrected for
  detector dead time and have had the background count rate and stim pulse
  rate subtracted. Data obtained when the Alice aperture door was closed are
  colored black; data obtained when the aperture door was open are colored
  red. The count rate produced by FUV photons during the door open periods was
  estimated by comparing the count rate immediately before/after the aperture
  door opened/closed and subtracted from the data.}
\end{figure}

\begin{figure}
\noindent\includegraphics[width=39pc]{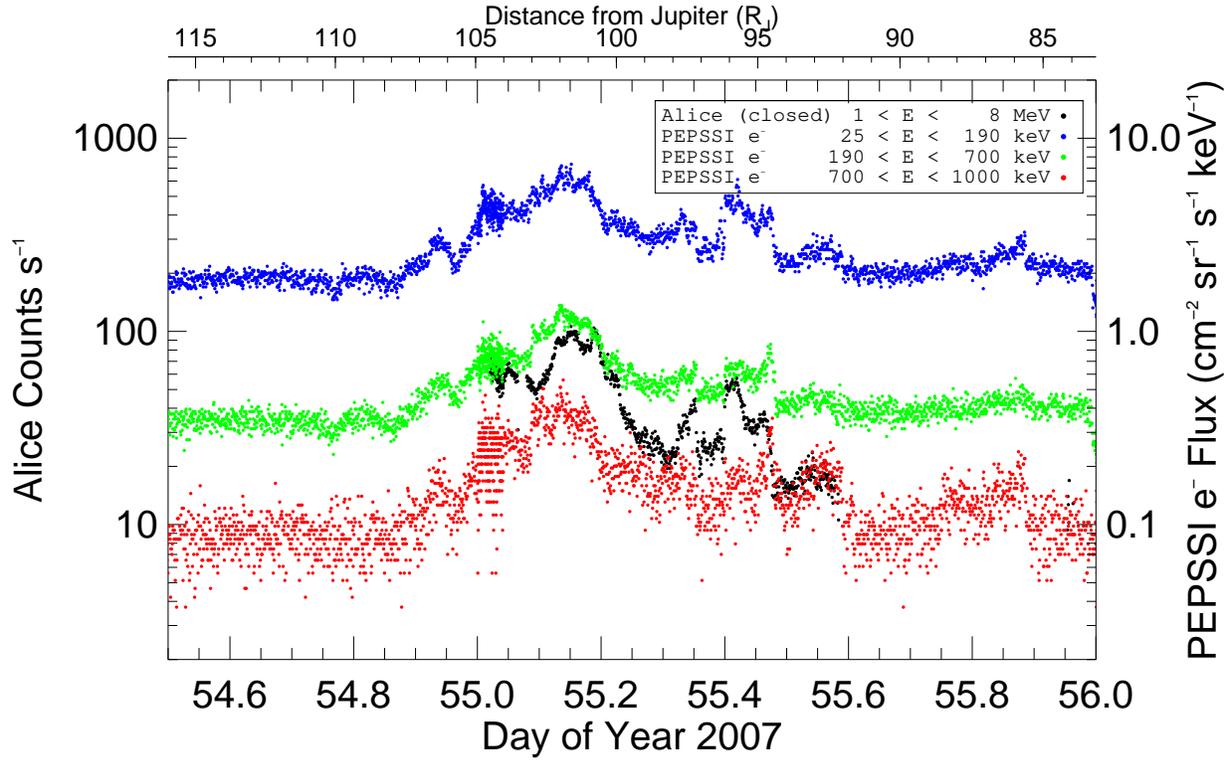}
\caption{\label{upstream_fig} Dead time corrected, background subtracted Alice
  count rates and PEPSSI electron fluxes, on day 55. PEPSSI electron data have
  been averaged over the three electron detectors.  During this period, New
  Horizons was upstream of the Jovian magnetosphere and immersed in the solar
  wind plasma.  Alice count rates have been averaged to match the 60-second
  sampling of the PEPSSI data during this period. Electron fluxes for PEPSSI's
  three energy ranges are shown in red, green, and blue.}
\end{figure}

\begin{figure}
\noindent\includegraphics[width=39pc]{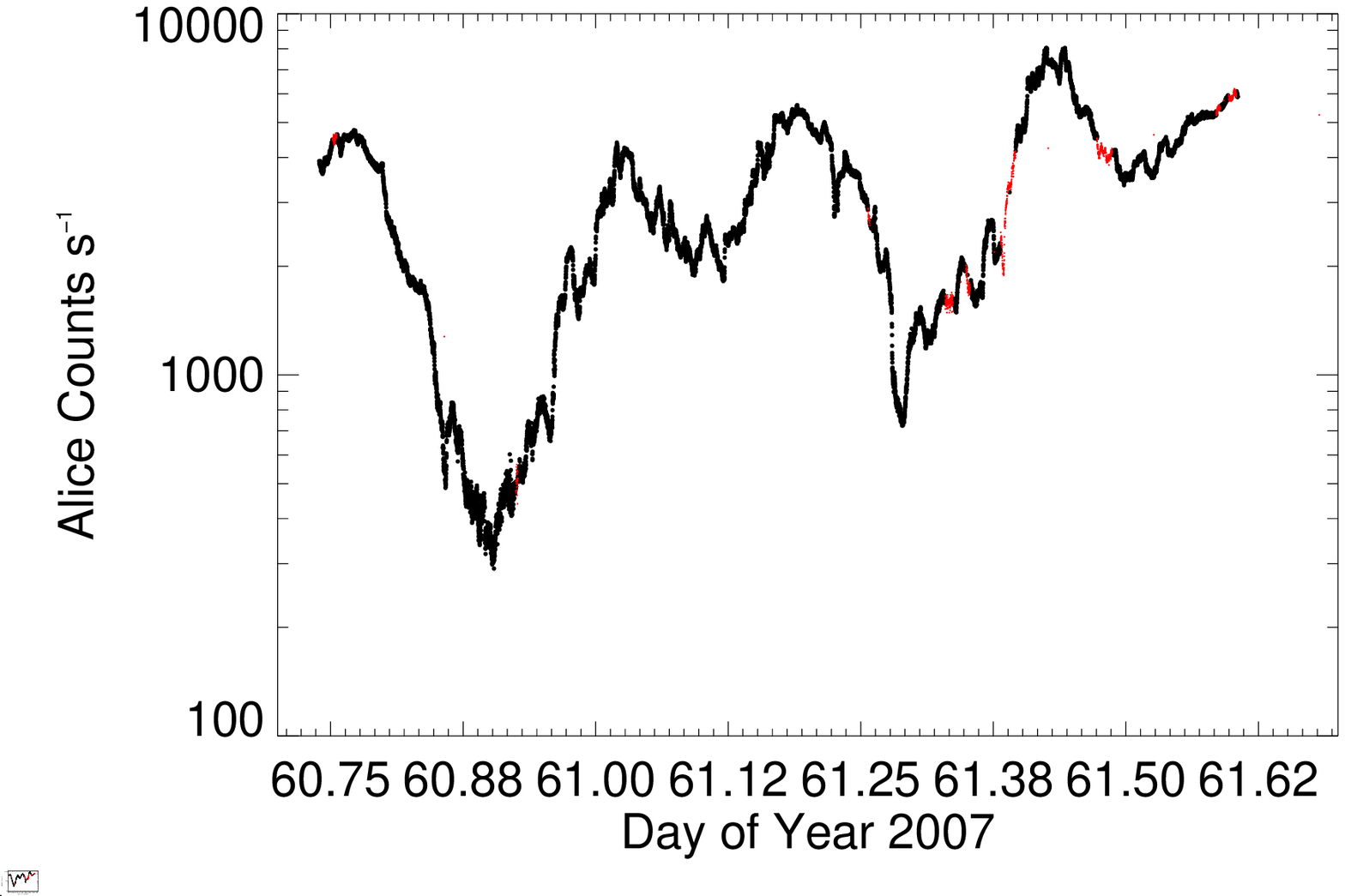}
\caption{\label{mpcross_fig} Dead time corrected, background subtracted Alice
  count rates and PEPSSI electron fluxes around the time of the Jovian
  magnetopause crossing.  The good match between Alice data obtained with the
  aperture door opened (black) and closed (orange) demonstrates the validity
  of the technique used to subtract the FUV photon count rate described in
  Section~\ref{fuv_correction}.  Alice count rates have been averaged to match
  the 60-second sampling of the PEPSSI data during this period. Electron
  fluxes for PEPSSI's three energy ranges are shown in red, green, and blue. }
\end{figure}

\begin{figure}
\noindent\includegraphics[width=20pc]{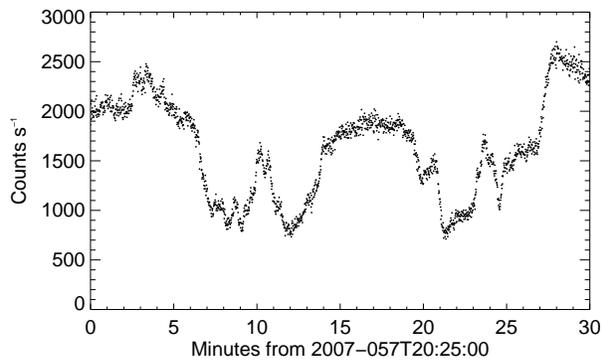}
\caption{\label{dayside_cnt_fig} Dead time corrected, background subtracted
  Alice count rate during a 30-minute period showing factor of two changes in
  count rate on timescales of a few minutes. During these observations, {\it
    New Horizons} was located 46.3~R$_J$ from Jupiter at approximately 1400
  LT.}
\end{figure}


\begin{figure}
\noindent\includegraphics[width=39pc]{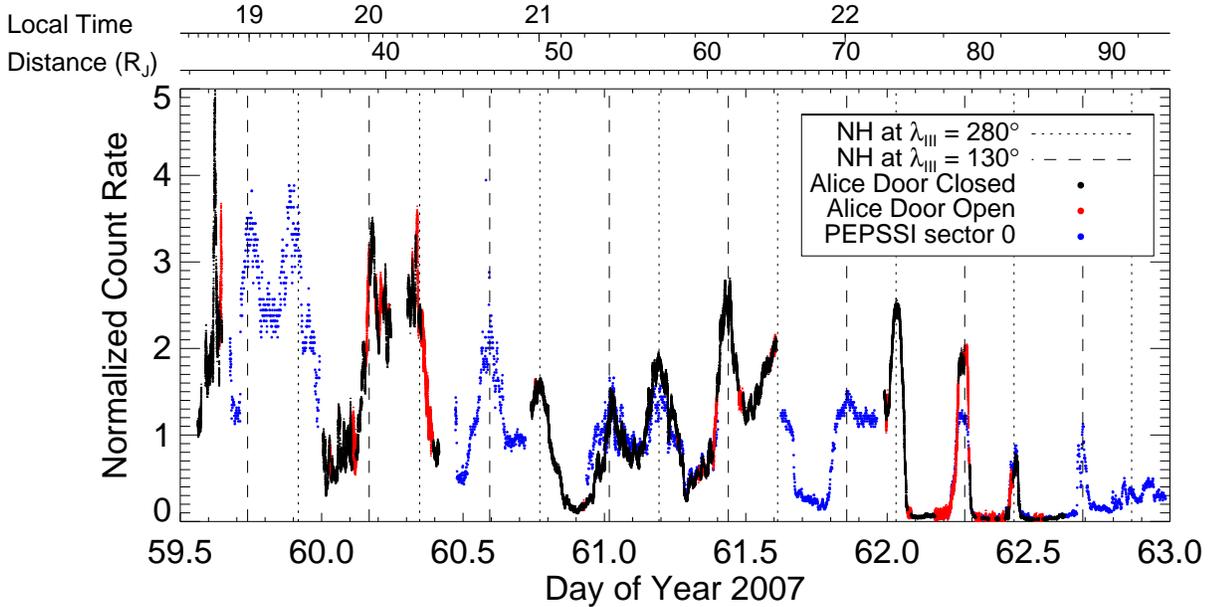}
\caption{\label{evening_cntrate_fig} Count rates for Alice and the PEPSSI E0
  700-1000~keV energy range. Count rates have been normalized to the average
  between DOY 61.0 and 61.4. The absolute count rate of Alice is approximately
  170 times greater than the absolute count rate in the PEPSSI E0 700-1000~keV
  energy range. The two axes above the plot show the local time of New
  Horizons, and the (spherical) radial distance from Jupiter in R$_J$. Times
  when the {\it New Horizons} spacecraft was located at $\lambda_{III}$ =
  130$^\circ$ and $\lambda_{III}$ = 280$^\circ$ are marked by dashed and
  dotted lines, respectively. Black dots represent count rates obtained when
  the Alice aperture door was closed and no FUV photons could reach the
  detector. Red dots represent the Alice count rate when the aperture door was
  open, after correcting for counts due to FUV photons.  The energetic
  electron flux at {\it New Horizons} shows a clear, double-peaked 10-hr
  periodicity corresponding to the Jovian current sheet sweeping over the
  spacecraft twice per Jovian rotation.}
\end{figure}

\begin{figure}
\noindent\includegraphics[width=20pc]{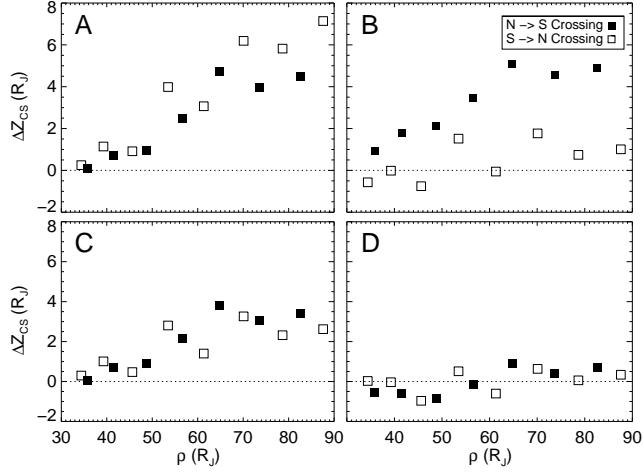}
\caption{\label{delta_z_fig} Residuals between the height of {\it New
    Horizons} above the rotational equator during the current sheet crossings
  listed in Table~\ref{cs_table} and the height of model current
  sheets. Filled symbols represent N$\rightarrow$S crossings, and open symbols
  represent S$\rightarrow$N crossings.  Panel A shows the residuals for a
  current sheet located in the magnetic dipole equator plane. Panel B shows
  residuals for the current sheet model of \cite{Khurana:schwarzl05}. Panel C
  shows residuals using the KS05 model and a value of $b_0=12.0^\circ$, which
  minimizes the difference between N$\rightarrow$S and S$\rightarrow$N
  crossings. Panel D shows the difference using the KS05 model with
  $b_0=12.5^\circ$ and a characteristic hinging distance, $x_H = 5$~R$_J$.}
\end{figure}


\end{document}